\documentclass[aps,prl,twocolumn,showpacs,superscriptaddress]{revtex4-1}

\usepackage{blindtext}
\usepackage{bm}
\usepackage{amsmath,amssymb,amsfonts,latexsym,fancyhdr,graphicx,epstopdf,times,txfonts}

\usepackage[english]{babel}
\usepackage{graphicx}
\usepackage{overpic}
\usepackage{xcolor}
\usepackage{color}
\usepackage{verbatim}
\usepackage{lipsum}
\usepackage{hyperref}
\usepackage{bbm}

\definecolor{redred}{HTML}{D53E4F}

\definecolor{greengreen}{HTML}{2F9B31}

\definecolor{blueblue}{HTML}{0059FF}

\newcommand{\wirr}{W_{\text{IRR}}}

\def\bra#1{\mathinner{\langle{#1}|}}
\def\ket#1{\mathinner{|{#1}\rangle}}

\def\bra#1{\mathinner{\langle{#1}|}}
\def\ket#1{\mathinner{|{#1}\rangle}}

\begin{document}

\title{Non-equilibrium quantum thermodynamics in Coulomb crystals}

\author{F. Cosco}
\affiliation{Turku Centre for Quantum Physics, Department of Physics and Astronomy, University of Turku, FI-20014 Turun yliopisto, Finland}
\author{M. Borrelli}
\affiliation{Turku Centre for Quantum Physics, Department of Physics and Astronomy, University of Turku, FI-20014 Turun yliopisto, Finland}
\author{P. Silvi}
\affiliation{Institute for Complex Quantum Systems and Center for Integrated Quantum Science and Technologies, Universitat Ulm, D-89069 Ulm, Germany}
\affiliation{Institute for Theoretical Physics, University of Innsbruck, A-6020 Innsbruck, Austria}
\author{S. Maniscalco}
\affiliation{Turku Centre for Quantum Physics, Department of Physics and Astronomy, University of Turku, FI-20014 Turun yliopisto, Finland}
\affiliation{Center for Quantum Engineering, Department of Applied Physics, Aalto University School of Science, P.O. Box 11000, FIN-00076 Aalto, Finland}

\author{G. De Chiara}
\affiliation{Centre  for  Theoretical  Atomic,  Molecular  and  Optical  Physics, Queen's  University  Belfast,  Belfast  BT7 1NN,  United  Kingdom}

\selectlanguage{english}

\begin{abstract}
We present an in-depth study of the non-equilibrium statistics of the irreversible work
produced during sudden quenches in proximity to the structural linear-zigzag transition of ion Coulomb crystals in 1+1 dimensions.
By employing both an analytical approach based on a harmonic expansion 
and numerical simulations, we show the divergence of the average irreversible work in proximity to the transition. We show that the non-analytic behaviour of the work fluctuations can be characterized in terms of the critical exponents of the quantum Ising chain.  Due to the technological advancements in trapped ion experiments, our results can be readily verified. 
\end{abstract}

\maketitle
\section{Introduction} 
The recently renovated interest in non-equilibrium thermodynamics of quantum systems, spurred from tremendous advances in experimental techniques, has found a plethora of interesting developments and applications \cite{CampisiRMP,EspositoRMP,GooldReview,AndersReview,XuerebReview}. From a theoretical perspective,  several important achievements are already available in the literature.
Prominent examples are the quantum generalisation of fluctuation relations such as the celebrated Jarzynski equality \cite{jarzynski1997nonequilibrium,tasaki2000jarzynski,crooks1999entropy,TalknerJPA2007,TalknerPRE2007} and the {design of a single-atom thermal machine \cite{AbahPRL2012}, recently realised with a trapped ion} \cite{RossnagelScience}.

In recent years, much interest has been devoted to the analysis of the quantum work extracted from, or absorbed by, a quantum system. 
While many definitions of work in a quantum setting have been proposed, the most popular one in the literature, based on two-time measurements \cite{TalknerPRE2007,CampisiRMP},
fulfils the Jarzynski equality but not the first law of thermodynamics when the system exhibits energy coherences \cite{perarnau2016no}.
Furthermore, several proposals have been put forward to estimate
work in quantum systems without the need of realising energy projections. These include schemes based on Ramsey interferometry \cite{Mazzola2013,Dorner2013,CampisiNJP} experimentally realised in an NMR setting \cite{Batalhao}. Other methods employ the aid of auxiliary continuous quantum systems \cite{RoncagliaPRL,DeChiaraPaz,villa2017cavity}. Some others are based on population imbalance and coherence in bosonic Josephson junctions \cite{Lena}.

In the many-body scenario, fluctuations of the work and of its irreversible contribution have been calculated mainly for spin chains in proximity
to a quantum phase transition \cite{Silva, Dorner, Fusco, Bayat, Paganelli}. In this context it has been shown that all the moments of the work probability distribution are singular when the system is dynamically driven close to the phase transition \cite{Bayat, Paganelli}.

In this context, most studies so far have  been 
 limited to archetypal examples of strongly correlated systems in condensed matter~\cite{Masca,Sindona,Soti,Yulia,Russo,Marino,Bayocboc,Zhong,DelCampoEntropy2016}. However, testing such predictions on experimental platforms poses additional challenges. In fact,
the full probability distribution seems to be unaccessible to observation because of the complexity of energy projection and because Ramsey schemes would involve, unrealistically, an ancilla coupled to the whole many-body system.

In this work, we analyse the out-of-equilibrium quantum thermodynamics of a model specifically tailored to an experimental setup.
Precisely, we consider ion Coulomb crystals (ICC): Many-body quantum systems of cold atomic ions confined to highly anisotropic traps and mutually interacting via Coulomb repulsion \cite{Dubin}.
Here we estimate the statistics of the irreversible work production during sudden quenches in proximity to the phase transitions in such ICC.
Our results can be experimentally tested in current experiments by measuring, as we show in this work, the transverse displacement distribution of the ions positions.

At equilibrium, ICC exhibit different structural arrangements, generally depending on the spatial properties of the trapping potential~\cite{DubinPRL1993}. The structural transitions occurring between these configurations are typically phase transitions of the first order \cite{Piacente}.
Here, we focus on quasi one-dimensional arrangements obtained for strongly anisotropic traps, which exhibit a
transition from a linear to a zigzag configuration \cite{Fishman}.
In the limit of ultracold ions, quantum fluctuations become relevant and the linear-zigzag transition becomes a quantum phase transition of the Ising universality class 
in 1+1 dimensions, at the thermodynamical limit~\cite{Shimshoni1,Shimshoni2,Cormick,silvi2013full,FishmansilviIOS,SilviPRB,PodolClaquant}. The production of defects during a quick change of the trap anisotropy has been studied theoretically \cite{delCampoPRL,DeChiaraNJP,Nigmatullin,SilviKZ} and experimentally \cite{ulm2013observation,pyka2013topological}.

With this setup in mind we analyse the fluctuations of the work performed upon the ICC system by changing suddenly the transverse confinement frequency near the linear-zigzag transition point. We compare analytical results from calculations based on a harmonic expansion with numerical calculations based on the density matrix renormalization group algorithm (DMRG) \cite{DMRG1,DMRG2} in the matrix product state formalism \cite{Ostlundrommer,AgeofMPS}. We show that when approaching the critical point such fluctuations display a singularity, and they exhibit a universal scaling compatible with the quantum Ising model. 

\section{Linear-Zigzag model for ICC} Ions confined to a fully anisotropic 3D trap and interacting via repulsive Coulomb interaction undergo a structural phase transition in which their spatial geometry changes from a one-dimensional linear chain to a planar zigzag configuration \cite{Fishman}. The control parameter of such a transition is the frequency $\omega$ of the transverse harmonic trapping.
This transition is driven by a mechanical instability of the chain
that is associated with a soft mode at the boundary of the Brillouin zone 
 whose frequency vanishes at the critical transverse trapping frequency.
In the following paragraph we briefly review the analytical approximations needed to recast the linear-zigzag transition into a simple short-range model. For convenience, we restrict the motion of the ions to the $XY-$plane in which the zigzag structure develops. Here $X$ is the direction parallel to the trap axis and $Y$ is the direction perpendicular to the trap axis that emerges as a result of spontaneous symmetry breaking or because of a small anisotropy in the transverse confinement.

The strong repulsion between the ions makes them practically distinguishable particles, which in turn allows us to write a Hamiltonian in a first quantization
\begin{equation}
\begin{aligned}
H_0 =&\sum_{j=1}^{L}\left[\frac{P_{x,j}^{2}+P_{y,j}^{2}}{2M}+\frac{M\omega_0^{2}}{2}Y_{j}^{2}+V_{L}(X_{j})\right]+\\
&\frac{Q^{2}}{8\pi\epsilon_{0}}\sum_{i\ne j}\left[(X_{i}-X_{j})^{2}+(Y_{i}-Y_{j})^{2}\right]^{-1/2}
\end{aligned}
\label{eq:hamiltonian1}
\end{equation}
in which $Q$ is the ion charge, $M$ is the mass, $(X_{j},Y_{j})$ and $(P_{x,j},P_{y,j})$ are the position and momentum of the $j-$th ion, respectively, and
$V_{L}(x)$ is the longitudinal component of the confining potential.
The quantum nature of this model stems from the commutator $[X_{i},P_{x,j}] = [Y_{i},P_{y,j}] = i \hbar \delta_{i,j}$.
As  shown in \cite{Fishman,Shimshoni1}, when the chain is sufficiently
close to criticality, the longitudinal and transversal components of $H$ effectively decouple.
One can therefore
fix the average equilibrium positions of the ions along the longitudinal direction to $x_j = j a$, with $a$ the effective Wigner lattice spacing.
Afterwards, the transverse dynamics Hamiltonian $H_y$ can be Taylor-expanded at fourth order in the displacements $y_j$.
The resulting theory, still long-range, can then be recast into a short-range model through an expansion of the scattering matrix
of the harmonic modes, at second order in $\delta k$ around the soft mode ($\delta k = k - \pi/a$) \cite{SilviPRB}.
This mapping effectively simplifies the Hamiltonian of Eq.~\eqref{eq:hamiltonian1} into
\begin{equation}
{H}(\omega) \!=\! \frac{1}{2}\!\sum_{j=1}^{L}  \left[- g^{2}\frac{\partial^{2}}{\partial {y}_{j}^{2}}+({\omega}^{2}-h_{1}){y}_{j}^{2}+
h_2({y}_{j}+{y}_{j+1})^{2}+h_3 {y}_{j}^{4} \right],
\label{eq:hamiltonian2}
\end{equation} 
in which $g=\sqrt{\hbar^{2}/Ma^{2}E_{0}}$ plays the role of an effective Planck constant measuring the impact of quantum 
fluctuations \cite{silvi2013full,PodolClaquant}.
Here all quantities have been expressed in dimensionless scales, according to: $H=H_0/E_{0}$ with $E_{0}=Q^{2}/(4\pi\epsilon_{0}a)$,
$y_j = Y_j/a$, 
and $\omega=\omega_0/\sqrt{E_{0}/Ma^{2}}$. Finally,
$h_{1}=7\zeta(3)/2, h_{2}=\ln 2$, and $h_{3}=93\zeta(5)/8$, with
$\zeta$ being the Riemann function, are universal constants \cite{silvi2013full,SilviPRB}.
The Hamiltonian of Eq.~\eqref{eq:hamiltonian2} captures accurately the dynamics of
the linear-zigzag quantum phase transition, and the critical point, identified by transverse frequency $\omega = \omega_C(g)$, can be 
computed as a function of $g$, and it was estimated to scale as $\omega_C(g) \approx h_1 - 3 h_3 g | \ln g | / 2\pi+ \mathcal O(g)$ for small $g$ \cite{PodolClaquant}.
In what follows we are going to extensively study the non-equilibrium
statistics of the irreversible work generated after sudden changes of the transverse frequency from $\omega_i=\omega$ to $\omega_f$ such that $|\omega_i^2-\omega_f^2|=\Delta\omega$.
First, we are going to present analytical results obtained using an approximated harmonic version of Eq.~\eqref{eq:hamiltonian2} and then compare them with numerical simulations, based on the DMRG algorithm.

\section{Non-equilibrium quantum thermodynamics} 
In quantum mechanics work is not a quantum observable \cite{TalknerPRE2007} but a generalised measurement \cite{RoncagliaPRL}. As such, it is strongly affected by quantum fluctuations arising in the measurement process. 
The key figure in this respect is the probability distribution of the work generated when the system is subject to a time-dependent Hamiltonian, but is otherwise isolated by sources of heat or dissipation.

In this paradigm the time-dependent Hamiltonian $H_{i}=  H( \omega_{i})$ is controlled via the frequency $\omega$. In turn, $\omega$ is quenched in time according to a certain time-dependent protocol $ \omega(t)$, within the time window $[t_i, t_f]$ [and accordingly, $\omega_i = \omega(t_i)$ and $\omega_f = \omega(t_f)$].
 At the beginning of the protocol the system is assumed at equilibrium
in the Gibbs state $\rho_i=e^{-\beta   H ( \omega_i)}/\mathcal{Z}_i$, where $\mathcal Z_i = \mathrm{ Tr}[e^{-\beta   H ( \omega_i)} ]$
is the partition function, and the inverse temperature $\beta$ is also expressed in dimensionless units.
For later convenience we also define the final equilibrium partition function $\mathcal Z_f = \mathrm{ Tr}e^{-\beta   H ( \omega_f)}$. 
Two sets of energy measurements are then performed, the first prior to the protocol and corresponding to the eigenstates of $  H_{i}$, and the second one right after the protocol and corresponding to the eigenstates of $  H_{f}$. One can define the work distribution performed during the $H_{i}\to H_{f}$ transformation as
\begin{equation}
P_{F}(W)\equiv \sum_{n,\bar m}  p^0_n p^{t_{f}}_{\bar m|n} \delta [W-( \epsilon_{\bar m}-\epsilon_n)],
\label{distributionP}
\end{equation}
in which $\epsilon_n$ and $\epsilon_{\bar m}$  are the eigenvalues of the initial and final Hamiltonian respectively, $p^0_n=e^{-\beta\epsilon_{n}}/\mathcal{Z}$
is the initial probability distribution in the energy levels,
and $p^{t_{f}}_{\bar m|n}=|\bra{\epsilon_{\bar{m}}} U \ket{\epsilon_{n}}|^{2}$ 
is the transition probability for the system to evolve from the state $\ket{\epsilon_{n}}$ to   $\ket{\epsilon_{\bar{m}}}$, after the time evolution
$U = U(t_i \to t_f) = \mathcal{T}\!\exp \int_{t_i}^{t_f} - i H(t) dt $.

For a sudden quench ($U=\mathbbm{1}$), the average work is simply $\langle W \rangle={\rm Tr}[\rho_i( H(\omega_f)- H(\omega_i) )]$ while  the free energy difference is $\Delta F=-\beta^{-1}\ln(Z_f/Z_i)$. Because of the relation $\langle W \rangle\ge \Delta F$, we define the irreversible work as the extra work needed to perform the transformation:
$\wirr\equiv\langle W \rangle- \Delta F$.

\section{Harmonic approximation}  For small quantum fluctuations, corresponding to small values of $g$,
the dynamics of the Wigner crystal described by the Hamiltonian \eqref{eq:hamiltonian2} can be expressed in terms of small quantum displacements, coupled harmonically, around the classical equilibrium ion positions. In the linear phase the classical equilibrium positions are $  y_j=0$ while in the zigzag phase these are $  y_j=(-1)^j b/2$ where the zigzag width $b$ is determined by $ \omega$ \cite{Fishman}.
In this regime one can find the normal frequencies associated with a normal mode at momentum $k\in[-\pi,\pi]$ of the harmonic chain of oscillators. In the linear phase these read
\begin{equation}
\omega_k^2=g^2 \left [\omega^2-h_1+ 4 h_2 \cos^2 \frac k2\right ]
\label{dispersionL}
\end{equation}
 At zero temperature, this semiclassical model predicts a critical transverse frequency at $ \omega_{C}=\sqrt{h_1}$, for which the frequency of the soft-mode at $k=\pi$ and the quadratic term in Eq.~\eqref{eq:hamiltonian2} vanish. For $  \omega> \omega_{C}$ the chain spatial
configuration is linear, while it is zig-zag in the opposite case. 
\begin{figure}[!t]
\centering
 \begin{overpic}[width = \columnwidth, unit=1pt]{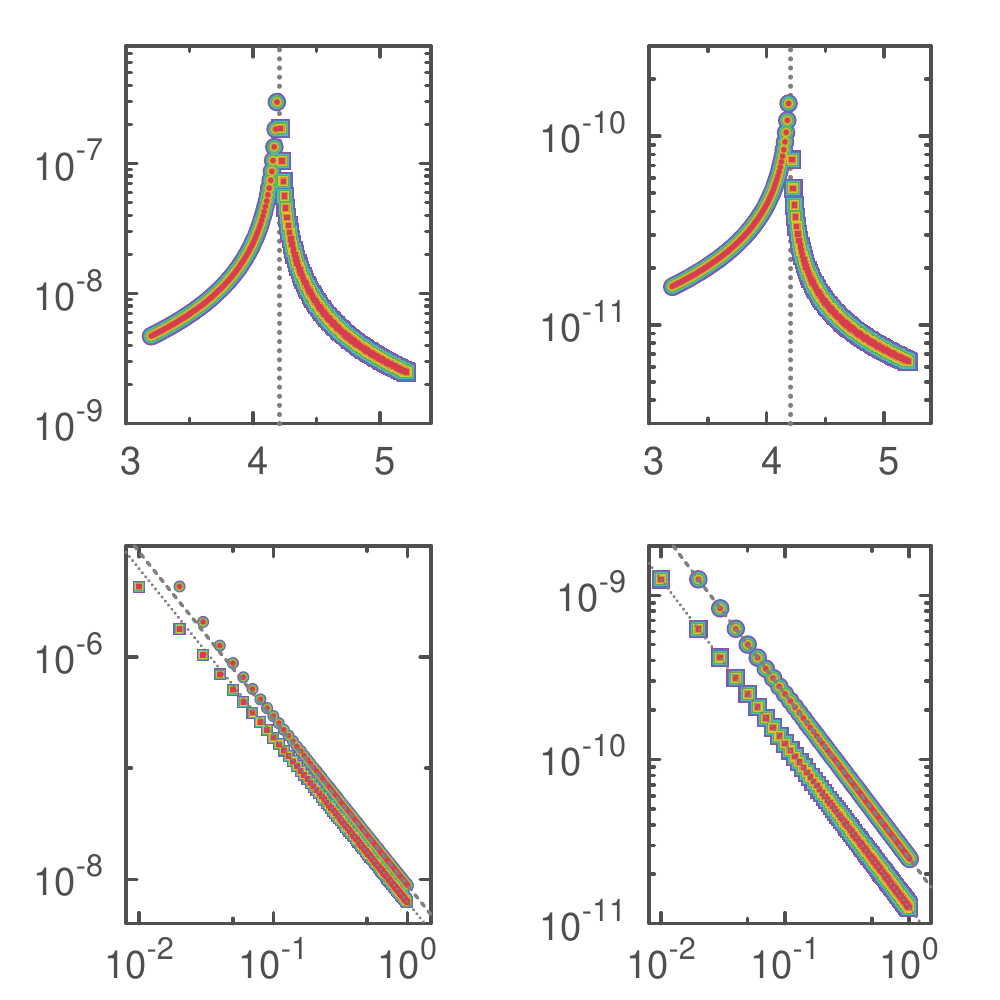}
 \put(56,93){$\displaystyle{\frac{\sigma^2_W}{L}}$}
 \put(82, 50){$\omega^2$}
 \put(86, 90){(b)}
  \put(2, 93){$\displaystyle{\frac{\wirr}{L}}$}
 \put(30, 50){$\omega^2$}
 \put(35, 90){(a)}
  \put(2, 40){$\displaystyle{\wirr^{\rm soft}}$}
 \put(30, -1){$|\omega^2-h_1|$}
 \put(35, 40){(c)}
  \put(56, 33){$\displaystyle{\sigma^2_{\rm soft}}$}
 \put(80, -1){$|\omega^2-h_1|$}
 \put(86, 40){(d)}
\end{overpic}
\caption{(color online) Upper panels: Irreversible work $\wirr$ (a) and the variance $\sigma^{2}$ (b) for small quenches with amplitude $\Delta\omega=0.01$ within the same phase (circles in the zigzag phase and squares in the linear phase) at zero temperature as a function of the initial frequency squared $\omega^2$. The vertical dotted line indicates the critical frequency.
Lower panels: Contribution of the soft mode to the irreversible work $\wirr$ (c) and the variance $\sigma^{2}$ (d) as a function of $|\omega^2-h_1|$.
The straight lines are the approximate scaling expressions \eqref{eq:Wirrscaling} and \eqref{eq:sigmascaling}.
Various system sizes $L$ have been considered here: $L=60$ (purple), 72 (blue), 90 (cyan), 108 (green), 120 (yellow), 132 (orange), and 144 (red).
} 
\label{wsigmaharmonic}
\end{figure}
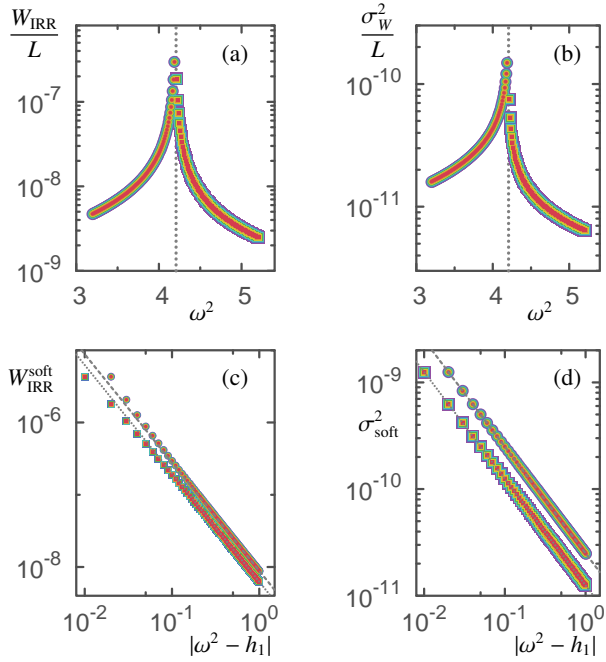

Instead of calculating directly the work probability distribution of Eq.~\eqref{distributionP}, we analytically compute its Fourier transform, namely the characteristic or moment-generating function:
\begin{equation}
\chi_{F}(t)\equiv \int dW e^{{iWt}{ }} P_F (W) =\mathrm{ Tr} [ e^{ {i H _f t}{ }} U(t_{f},0)^\dagger   e^{-{i H_i t}{ }}  U(t_{f},0)\rho_i]
\end{equation}
similarly to the methods reported in \cite {DeffnerLutz,Carlisle}. We extract 
the average irreversible work $\wirr$ and its statistical variance $\sigma^2_{W}=\langle W^2\rangle - \langle W\rangle^2$ by computing the first two moments of $\chi_{F}(t)$.
For an instantaneous quench within the same phase one finds
\begin{equation}
\begin{aligned}
\wirr=\sum_k  \Bigg[\frac {1}{2}  \left(\Omega_k\omega_{k}^{f}-\omega_{k}^{i}\right) \coth \frac{\beta \omega_k^{i}}{2}
-\frac {1}{\beta}\ln \frac {\sinh {(\frac{\beta \omega_k^{f}} {2})}}{\sinh ( \frac {\beta  \omega_k^{i}}{2})}\Bigg],
\end{aligned}
\label{wsamephase}
\end{equation}
\begin{equation}
\sigma^2_W=\sum_k \frac{{\omega_k^{f}}^2 \cosh (\beta  {\omega_k^{i}}) \left(\Omega_k^2-1\right)+\left({\omega_k^{f}}\Omega_k^2-{\omega_k^{i}}\right)^2}{4 \sinh ^2( \beta  {\omega_k^{i}}/2)},
\label{sigmasamephase}
\end{equation}
in which $\omega_{k}^{i(f)}$ is the initial (final) frequency of the $k$ mode and $\Omega_{k}=(\omega_k^{i2}+\omega_k^{f2})/2\omega_k^{i}\omega_k^{f}$. The irreversible work and the work variance, at zero temperature, are shown in Fig.~\ref{wsigmaharmonic} for small quenches within the same phase and for chain lengths ranging from 60 to 144 ions. It is interesting to note the extensiveness of both the irreversible work and its variance. In agreement with previous results \cite{Silva,Dorner,Fusco,Bayat} both $\wirr$ and $\sigma^2_W$ diverge at the critical frequency as a consequence of the vanishing of the lowest eigenfrequency.

While the harmonic approximation works well far from the critical point, the vanishing soft mode frequency causes an unphysical divergence, even for a finite number of ions, as the chain approaches criticality, as evidenced in Fig.~\ref{wsigmaharmonic}.  On both the linear and the zigzag sides of the transition, the irreversible work and the statistical variance are monotonically increasing functions for $\omega\to\omega_C$ respectively.  Due to the vanishing excitation gap at $k=\pi$ (soft mode), any quench  close to the critical point, no matter how small, will always require an amount of work much larger than the mere energy difference between the two equilibrium configurations: $\langle W\rangle \gg\Delta F$.

 In order to understand better this divergence, we isolate the contribution to the irreversible work and to the variance of the soft mode. Limiting the sums in Eqs.~\eqref{wsamephase} and \eqref{sigmasamephase} to $k=\pi$ and expanding up to second order in $\Delta\omega^2$ we obtain
 \begin{eqnarray}
 \label{eq:Wirrscaling}
\wirr^{\rm soft} &=&\frac{g \Delta \omega ^2}{8\gamma_W \left|\omega ^2-h_1\right| ^{3/2}}+\mathcal O(\Delta\omega^3)
\\
\label{eq:sigmascaling}
\sigma_{\rm soft}^{2}&=&\frac{g^2\Delta \omega ^2 }{\gamma_\sigma |\omega ^2-h_1|}+\mathcal O(\Delta\omega^3)
\end{eqnarray}
Both expressions are valid in the linear phase by taking $\gamma_W=2$ and $\gamma_\sigma=8$ and in the zigzag phase by taking $\gamma_W=\sqrt 2$ and $\gamma_\sigma=4$. The expression for $\sigma_{\rm soft}^{2}$ is exact in the linear phase. These results are shown in the inset of Fig.~\ref{wsigmaharmonic}. 

\begin{figure}
 \begin{center}
 \begin{overpic}[width = \columnwidth, unit=1pt]{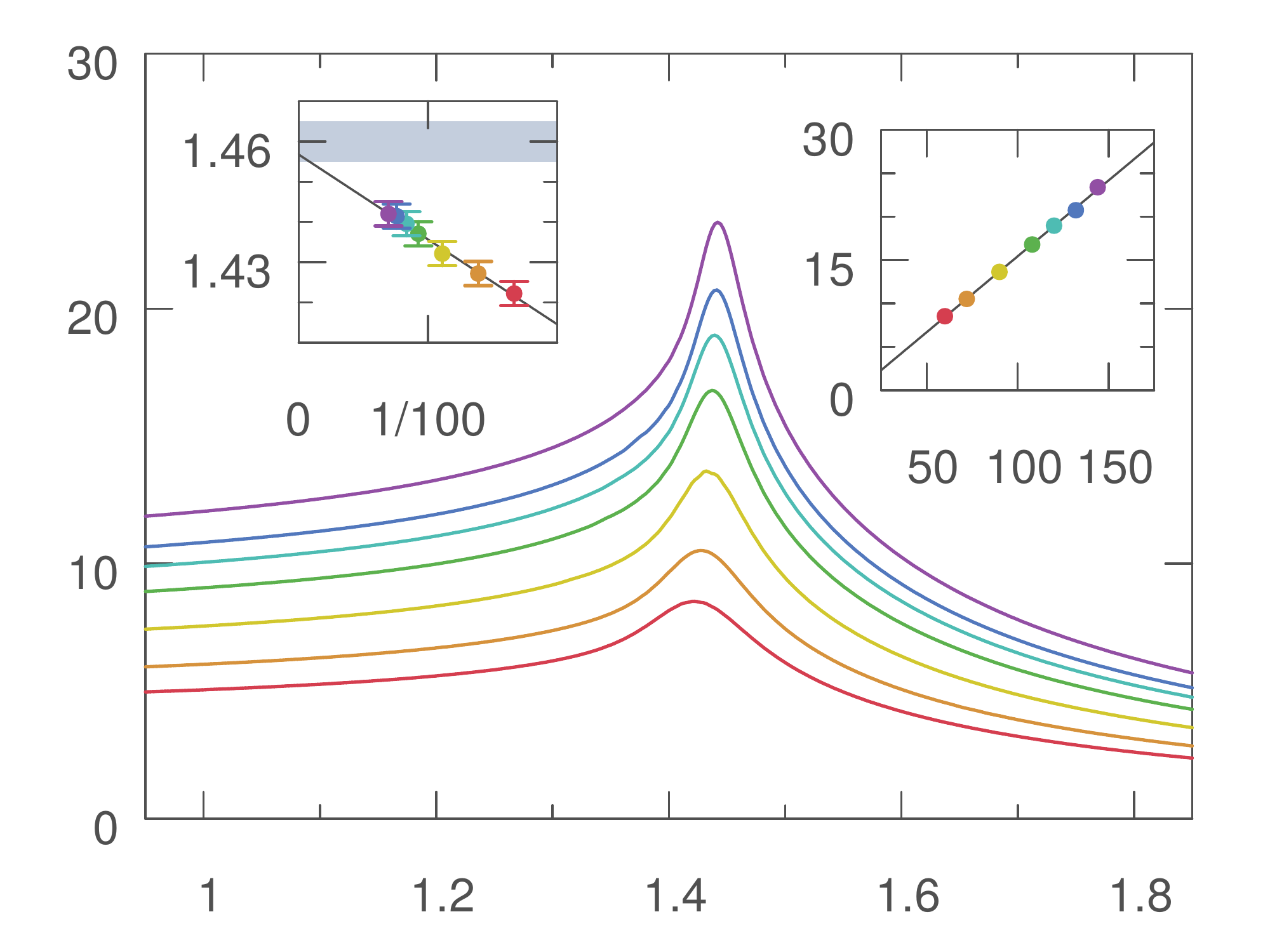}
 \put(10, 72){$\times 10^{-6}$}
 \put(0, 60){$\wirr$}
 \put(80, 0){$\omega^{2}$}
 \put(14, 58){$\omega_{\star}^{2}$}
 \put(45, 63){$\omega_{C}^{2}$}
 \put(41, 43){$1/L$}
  \put(59, 58){$\wirr^{(max)}$}
 \put(68, 66){$\times 10^{-6}$}
 \put(84, 33){$L$} 
 \end{overpic}
\end{center}
\caption{ \label{fig:two} (color online) 
Main panel: Irreversible work $\wirr $ at $T=0$ for small quenches in $\omega^{2}$, considering and increasing number $L$ of ions. Here
$L=60$ (red), 72 (orange), 90 (yellow), $108$ (green), $120$ (cyan), $132$ (blue) and $144$ (purple).
Left inset: Location $\omega_{\star}^2$ of the maximum of $\wirr$ in the control parameter $\omega^2$, as a function of the system size $L$.
Extrapolation to the thermodynamical limit of $\omega_{\star}^2$, performed via fitting (gray line), shows that it is compatible with the critical point
$\omega_C^2$ calculated via the DMRG algorithm (shaded area).
Right inset: Value of the maximum of $\wirr$ as a function of the ion chain length $L$, showing an accurate linearity.
}
\end{figure}
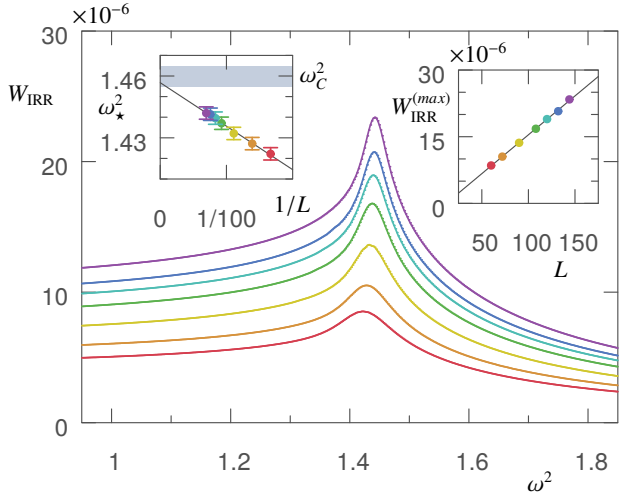

The scaling found for the soft mode is however modified when adding the other modes. In the thermodynamic limit and close to criticality we can expand Eq.~\eqref{wsamephase} in the linear phase, at $T=0$, as
\begin{eqnarray}
\wirr(T=0) &=&\sum_k \frac {(\omega_k^f-\omega_k^i )^2}{4 \omega_k^i}  \simeq L \int dk \, \frac {(\omega_k^f-\omega_k^i )^2}{4 \omega_k^i}
\nonumber \\
& \sim&  L\ln\left(\omega^2-h_1\right)+A,
\label{wirranalytic}
\end{eqnarray}
in which the constant $A$ depends on a small momentum cut-off introduced when turning the summation in Eq.~\eqref{wsamephase} into an integral. A similar expression holds in the zigzag phase \cite{RamseyDeChiara}.

We remark that the results of this section have been obtained assuming the short range effective model of Eq.~\eqref{eq:hamiltonian2}. We obtain similar expressions, within the harmonic approximation, for the long-range model of Eq.~\eqref{eq:hamiltonian1} finding the same scaling with renormalised parameters, e.g. the critical frequency is modified and has a weak finite-size correction $\sim 1/L^2$. So far, however, we have neglected the non-linear couplings between the normal modes. To overcome this,  in the next section we solve numerically the full anharmonic problem.

\section{Full anharmonic model}
In this section we present numerical results from the treatment of the full short-range Hamiltonian~\eqref{eq:hamiltonian2}. These results rely on the assumption that the initial
state is prepared at zero temperature $\beta \to \infty$, and thus can be found via variational methods.
In fact, the quantum many-body ground states $|\Psi_G(\omega)\rangle$ of the Hamiltonian from Eq.~ \eqref{eq:hamiltonian2} are simulated with the DMRG algorithm,
using a numerical technique for continuous-variables quantum systems analogous to Refs.~\cite{continuousvariableMPS,silvi2013full,SilviPRB}.
For any given $\omega$, we evaluate the corresponding ground state energy
$E_G(\omega) = \langle \Psi_G(\omega) | H(\omega) |\Psi_G(\omega)\rangle$ and the total fluctuation of the transverse displacement operators
$\mathcal{Y}^2(\omega) = \sum_j \langle \Psi_G(\omega) | y^2_j |\Psi_G(\omega)\rangle$.
Such data are sufficient to evaluate the average work $\langle W \rangle$  arising from a sudden quench
where $\omega$ is instantaneously driven from $\omega_i$ to $\omega_f$. In fact, we can simplify
\begin{eqnarray}
\langle W \rangle &=& \langle \Psi_G(\omega_i) | \left[H(\omega_f) - H(\omega_i) \right ]|\Psi_G(\omega_i)\rangle= \frac 12 \Delta\omega
\mathcal{Y}^2(\omega)\nonumber
\\
\end{eqnarray}
which we can easily compute from the equilibrium data we acquired.
Moreover, this expression for the average work shows how to measure it in an experiment by estimating the average quadratic transverse displacement of the ions.

Fig.~\ref{fig:two} displays the irreversible work generated by a small quench of the Hamiltonian \eqref{eq:hamiltonian2}, at $T=0$ and for several values of $L$.
The first feature we notice is the disappearance of the divergence at the critical point.
This is a clear signature of the finite-size effects in the quantum many-body system: at finite size $L$ the energy gap remains finite for all $\omega$,
thus actually smearing out the non-analyticity. On the other hand, the harmonic approach exhibits a critical behaviour even at finite size.
Moreover,  by increasing the number of ions two features appear: the peak in $\wirr$ becomes increasingly
sharper and its position slowly shifts towards larger frequencies, in contrast to the harmonic theory.

As for the value of the peak itself, we expect it to grow linearly in $L$,  $\wirr$ being an extensive
quantity and as already derived from the harmonic theory, see Eq.~\eqref{wirranalytic}. Fig.~\ref{fig:two} fully confirms this prediction.
In this case we  can also draw a direct  connection between our findings and recently results in the study of infinitesimal quenches and ground state fidelity susceptibility for the quantum Ising model \cite {Paganelli}. In this respect, one can investigate the finite-size effect on the maxima positions by plotting $\omega_{\textrm{max}}$ as a function of $1/L$, see Fig.~\ref{fig:two}, left inset. 

To gain further insight into the behaviour of the irreversible work, we adopt the finite-size scaling ansatz 
with a model-dependent scaling function $f$
\begin {equation}
1-e^{\frac {\wirr-\wirr^{\rm(max)}}{L}}=f\left[L^{\frac{1}{\nu}}(\omega^2-\omega^2_\star)\right]
\label{scaleansatz}
\end{equation}
which is typical of quantities that diverges logarithmically with the control parameter as in the case of the specific heat in the 2D classical Ising model \cite{Barber}.

\begin{figure}[t]
 \begin{center}
 \begin{overpic}[width = \columnwidth, unit=1pt]{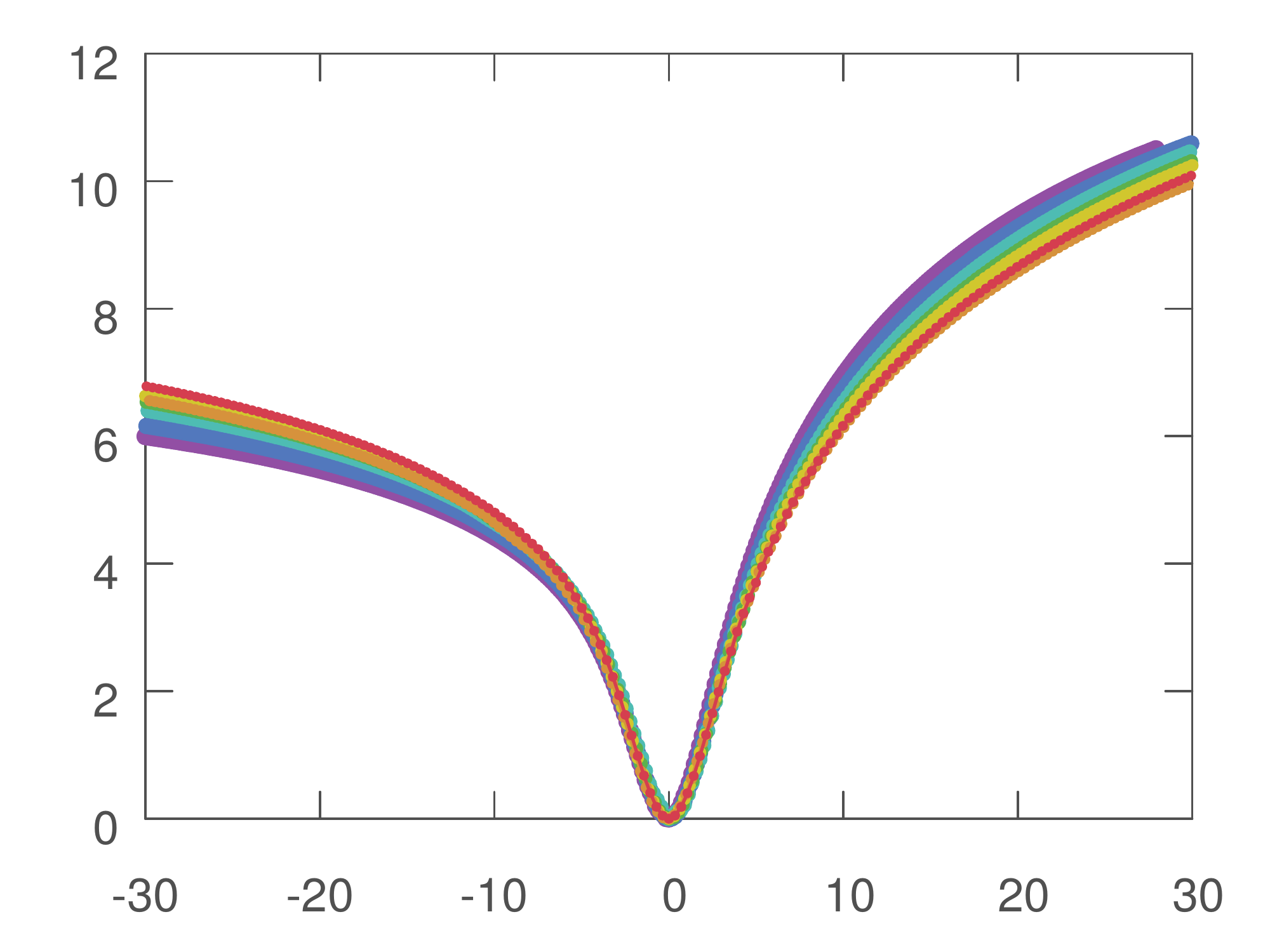}
 \put(10, 72){$\times 10^{-8}$}
 \put(-2, 40){\rotatebox{90}{$1-\exp [{\frac {\wirr-\wirr^{\rm (max)}}{L}}]$}}
 \put(75, -2){$L(\omega^2-\omega^2_\star)$}
 \end{overpic}
\end{center}
\caption{ \label{fig:three} (color online) 
Collapse of the irreversible work of the full anharmonic model according to the scaling ansatz defined in Eq.~ \eqref{scaleansatz}. The color coding for the symbols is as in Fig.~{fig:two}.}
\end{figure}

In Fig.~\ref{fig:three} we show the rescaled data and we obtain a collapse of the irreversible work as in the Ising model with the same critical exponent $\nu=1$. We remark that given the magnitude of the data shown in Fig.~\ref{fig:three}, a similar collapse plot (not shown) is obtained by expanding the exponential function: $1-\exp[(\wirr-\wirr^{\rm(max)})/L]\simeq (\wirr^{\rm(max)}-\wirr)/L$.

\section{Conclusions} In this work we have investigated the irreversible work production associated with infinitesimal quenches of the transverse frequency around a structural phase transition of Coulomb crystals. We have employed different approaches starting from a harmonic approximation that allows us to obtain analytical results but in turn generates an unphysical vanishing gap at $k=\pi$ at the critical point, which is known to drive the structural phase transition. We isolated the contribution of the soft mode to the irreversible work finding a power law scaling of the irreversible work. Finally, we have studied the full anharmonic model through DMRG $T=0$ simulations. With these different approaches we have observed the extensiveness of the irreversible work, and how they generate different scaling properties. Interestingly, the scaling laws recently found for the irreversible work of the Ising model, in terms critical exponents and collapse ansatz, are recovered in the full anaharmonic model when the $g$ parameter is big enough to appreciate shifts from the classical critical point due to pure quantum effects.  
Beyond the fundamental interest, it has been found that such critical behaviours could be used to design a quantum Otto engine where it has been showed that, for a working substance around criticality, the Carnot point can be reached \cite {Campisi:2016aa}. In this sense our setup is of particular experimental interest, since we have also shown how the calculated quantities can be related to the fluctuations in the transverse displacement of the ions.

\acknowledgements
We acknowledge support from the Horizon 2020 EU collaborative projects QuProCS (Grant Agreement 641277) and
TherMiQ (Grant Agreement 618074). PS gratefully acknowledges support from the EU via UQUAM and RYSQ, the DFG via SFB/TRR 21, and the Baden-WŸrttemberg Stiftung via Eliteprogramm for Postdocs.

\bibliographystyle{apsrev4-1}
\bibliography{Wirr_biblio}

\end{document}